\RequirePackage{fix-cm}
\documentclass[twocolumn]{svjour3}          
\smartqed  
\usepackage{graphicx}
\usepackage{color}
\newcommand {\cb} {\color{blue}}
\begin{document}

\title{Multigap Superconductivity in GdFeAsO$_{0.88}$ Evidenced by SnS-Andreev Spectroscopy}


\author{
T.E.~Shanygina \and
S.A.~Kuzmichev \and
M.G.~Mikheev \and
Ya.G.~Ponomarev \and
S.N.~Tchesnokov \and
Yu.F.~Eltsev \and
V.M.~Pudalov \and
A.V.~Sadakov \and
A.S.~Usol'tsev \and
E.P.~Khlybov \and
L.F.~Kulikova
}


\institute{T.E.~Shanygina \and A.S.~Usol'tsev \at
Lebedev Physical Institute RAS, 119991 Moscow, Russia \\
Lomonosov Moscow State University, 119991 Moscow, Russia\\
Tel.: +7(499)1326780\\
Fax: +7(499)135-78-80\\
\email{tatiana.shanygina@gmail.com}
\and
S.A. Kuzmichev \and M.G.~Mikheev \and Ya.G.~Ponomarev \and S.N.~Tchesnokov \at
Lomonosov Moscow State University, 119991 Moscow, Russia
\and
Yu.F.~Eltsev \and V.M.~Pudalov \and A.V.~Sadakov \at
Lebedev Physical Institute RAS, 119991 Moscow, Russia
\and
E.P.~Khlybov \and L.F.~Kulikova \at
Institute for High Pressure Physics RAS, 142190 Troitsk, Russia
}

\date{Received: date / Accepted: date}

\maketitle

\begin{abstract}
Using intrinsic multiple Andreev reflection effect (IMARE) spectroscopy we studied
superconducting properties of nearly optimal oxygen-deficient \\ GdFeAsO$_{0.88}$ polycrystalline samples (bulk critical temperatures $T_C^{bulk} = 49 \div 52$\,K). Temperature dependences for two superconducting gaps $\Delta_{L,S}(T)$ ($T_C^{local} = 48 \div 50$\,K) have been measured in the range from 4.2 to 50\,K.
The  $\Delta_{L,S}(T)$ dependences were found to deviate
from the BCS-like function; this suggests an importance of the $k$-space (internal) proximity effect between the two condensates.
\keywords{two-gap superconductivity \and 1111 pnictides \and SnS-Andreev spectroscopy \and "break-junction"}
\PACS{74.25.-q \and 74.45.+c \and 74.70.Xa}
\end{abstract}

GdFeAsO firstly reported in \cite{Cheng} belongs to oxypnictide high-temperature superconductors \cite{Kamihara}. The stoichiometric compound is an antiferromagnetic metal with spin-density-wave (SDW) ground state \cite{Klauss}, which can be turned into superconductivity under electron doping. The maximal critical temperatures up to $T_C = 56$\,K are achieved by O-deficiency, F introducing instead O \cite{Khlybov} or Gd substitution for Th \cite{Wang}. The Fermi surface for 1111 materials contains quasi-two-dimensional electron and hole sheets suggesting a possibility of multi-gap superconductivity \cite{Coldea}.

Our previous studies established the presence of two distinct superconducting gaps $\Delta_L = 10.5 \pm 2$\,meV and $\Delta_S = 2.3 \pm 0.4$\,meV in fluorine-doped GdFeAsO$_{1-x}$F$_{x}$ ($T_C^{bulk} = 53 \pm 1$\,K) \cite{Gd,UFN,GdJPCS}. Here we report temperature dependences of superconducting gaps $\Delta_{L,S}(T)$ in oxygen-deficient GdFeAsO$_{0.88}$ measured by intrinsic multiple Andreev reflections effect (IMARE) spectroscopy of array superconductor - normal metal - superconductor (SnS) contacts.

High-pressure synthesis of GdFeAsO$_{1-\delta}$ polycrystalline samples with bulk critical temperatures $T_C^{bulk} = 49 \div 52$\,K is detailed in \cite{Khlybov}. To realize the IMARE spectroscopy, superconductor - constriction - superconductor (ScS) contacts were formed
in the samples by the ``break-junction'' technique \cite{Moreland}. The samples (thin plates of $2 \times 1 \times 0.2$\,mm$^3$) were attached with two current and two potential leads to the spring holder by liquid In-Ga alloy. Bending the spring holder at $T = 4.2$\,K causes a microcrack generation into the sample, which ensures
studies of clean cryogenically cleaved superconducting surfaces separated with a constriction (ScS-contact). The ``break-junction'' technique
enables easy readjustment of a contact point, which
facilitates measuring tens of contacts during an experiment. While tuning of a break junction, it is important not to slide apart the two superconducting banks of the sample to avoid the cryogenic clefts degradation. Besides, a contact point is naturally located in the bulk of the sample thus excluding an overheating of the contact point.

$R(T)$-dependences  and current-voltage characteristics (CVC) of our ScS break junctions were measured with an NI digital board. The dynamic conductance $dI(V)/dV$ curves were measured by a standard current modulation technique \cite{LOFA,Rakhmanina}.

K\"{u}mmel et al. \cite{Kummel} showed the CVC of superconductor - normal metal - superconductor (SnS) contact to have an excess current at zero bias and a subharmonic gap structure (SGS) - series of dynamic conductance minima at bias voltages
\begin{equation}
V_n = \frac{2\Delta}{en},~~n = 1, 2,\dots
\end{equation}
due to the multiple Andreev reflections effect. Earlier \cite{JETPL04,SSC04}, the two
similar structures were observed for a two-gap superconductor. The Andreev conductance dips of a large order $n \geq 3$ become clearly visible only at dI/dV-curves of Sharvin-type \cite{Sharvin} SnS-contact (with diameter $a$ less than the quasiparticle mean free path $l$). The experimental $dI/dV$ are typical for classical SnS-contact with excess-current characteristic and, therefore, well-described by the model by K\"{u}mmel et al. \cite{Kummel}.

Layered structure of Gd-1111 also provides the presence of steps-and-terraces at cryogenic clefts. Such natural structures being typical for the ``break-junction'' technique can form stack contacts corresponding a sequence of individual ScS-contacts (junctions of S-c-S-c-\dots-S-type) and, therefore, provide observation of intrinsic multiple Andreev reflections effect (similar to intrinsic Josephson effect \cite{Nakamura}. This was observed for the first time in Bi-2201 \cite{Bi}). Obviously, bias voltage at $dI/dV$-curve of such an array scales with number $N$ of junctions in the stack (in comparison with a single contact $dI/dV$-characteristic).
Therefore, $dI/dV$-measurements for the stack contacts
reduce surface defects influence and
increase by a factor of $N$
 accuracy of gap value measurements \cite{Li}.

\begin{figure}
  \includegraphics[width=0.48\textwidth]{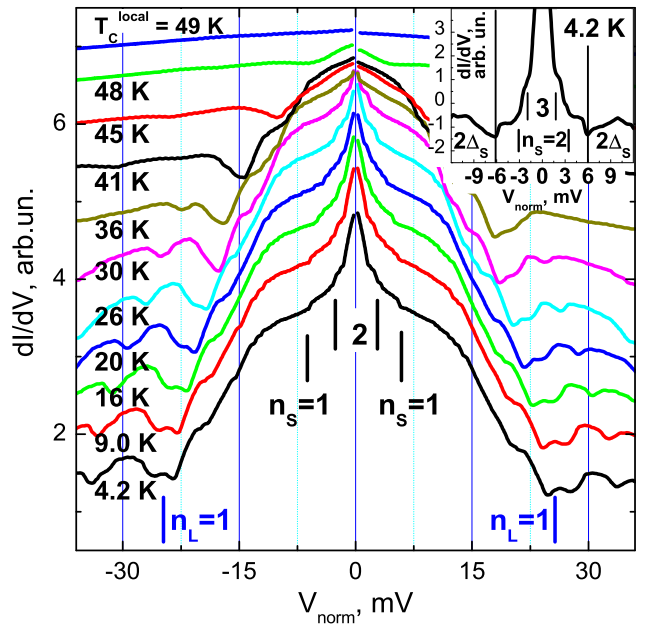}
\caption{Dynamic conductance of SnS-Andreev array (sample Khl8, contact $\sharp f$, 3 junctions in the stack) measured within the range 4.2\,K$\leq T \leq T_C^{local} = 49$\,K. Vertical lines mark the Andreev dips position for the large gap $\Delta_L \approx 12.5$\,meV ($n_L$ labels) and the small gap $\Delta_S \approx 3$\,meV ($n_S$ labels) at $T = 4.2$\,K. All the curves are normalized to a single junction and shifted along the vertical scale for the sake of clarity. \textbf{Inset}: The enlarged low-bias fragment of the dI/dV ($T = 4.2$\,K) containing the SGS of the small gap ($n_S$ labels). Background is suppressed.}
\end{figure}

For symmetrical SnS-contacts, gap value can be calculated from dynamic conductance dips position directly by formula (1) within the whole temperature range $T \leq T_C^{local}$ \cite{Kummel} (note, on the contrary, $dI(V)/dV$ of SN-contact are to be fitted using 7 parameters in a case of a two-gap superconductor \cite{BTK}). Fig.~1 shows the dynamic conductance of SnS-array (sample KHL8, contact $\sharp f$, 3 junctions in the stack) measured at 4.2\,K$\leq T \leq T_C^{local}$. The spectra were normalized to a single junction and shifted along the vertical scale for the sake of clarity. Andreev dips corresponding to superconducting gaps are marked at the lower curve ($T = 4.2$\,K). Minima $n_L = 1$ located at $V_1 = \pm 25$\,mV define the large gap value $\Delta_L \approx 12.5$\,meV. Doublet-shape of the minima well-distinguished up to 36\,K seemingly originates from the large gap order parameter anisotropy about 8\% $\Delta_L$ (the $n_L = 1$ position is marked at the middle of the doublet). A fine structure (additional peculiarities above and below the $V_1$) might signify Leggett mode \cite{Leggett,SSC04} and requires further studies. The SGS corresponding to the small gap becomes clear at the enlarged low-bias fragment of the $dI(V)/dV$ measured at $T = 4.2$\,K (background was suppressed, see the inset of Fig.~1). Minima positions at $V_n =$ (6, 3 and 2)\,meV (being the $n_S = 1,~2,~3$, correspondingly) yield the small gap value $\Delta_S \approx 3$\,meV. The peculiarities marked as $n_S = 1$ could not be
associated with the third Andreev reflexes for the large gap ($n_L = 3$), because no fundamental $n_L = 2$ minima are visible. As temperature increases, the $\Delta_L$ and $\Delta_S$ dips smear and shift toward the low bias voltages.
The value of the local critical temperature $T_C^{local} = 49$\,K is obtained from the $dI/dV$ linearization (the contact area transition to the normal state).

As was mentioned above, the SnS-Andreev spectroscopy allows one to get superconducting gaps temperature dependence
with no additional fitting. The $\Delta_{L,S}(T)$ dependences plotted using data of Fig.1 are presented in Fig.~2. The standard single-gap BCS-like functions are depicted for the comparison by dash-dotted lines. Referring to Fig.2, the $\Delta_S(T)$ dependence strongly deviates from the single-band curve: the steep fall is followed by slow fading, like a ``tail'', which tends to the $T_C^{local}$. Such the behavior is reflected at the $\Delta_L(T)$ dependence as a slight bend in
comparison with the single-band curve. Excepting the bend, the large gap temperature dependence complies with the theoretical curve. Thus, the two superconducting gaps approach zero at the common $T_C^{local}$. The inset of Fig.~2 compares aforementioned temperature dependences normalized to 1, $\Delta_{L,S}/\Delta_{L,S}(0)$, versus $T$. The difference in the gaps temperature course is obvious, which concludes the peculiarities observed at Fig.1 to reflect the properties of two distinct superconducting condensates.

\begin{figure}
  \includegraphics[width=0.48\textwidth]{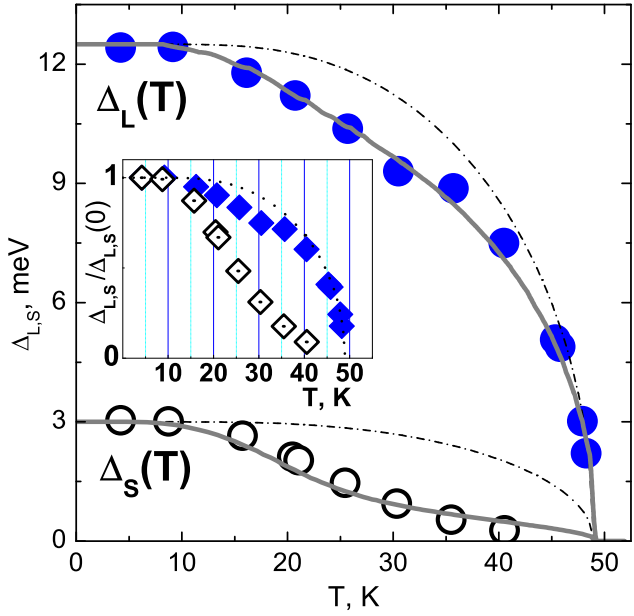}
\caption{The large (solid circles) and the small (open circles) gaps temperature dependence plotted using data of Fig.1. The curves are approximated by single-gap BCS-like function (dash-dotted lines) and two-band one (solid lines). \textbf{Inset}: The normalized gaps $\Delta_{L,S}/\Delta_{L,S}(0)$} temperature dependence (solid and open rhombs, correspondingly). Single-band BCS-like function is presented for comparison.
\end{figure}

The experimental $\Delta_{L,S}(T)$ dependences were successfully fitted
with two-gap BCS-like equations, established by Moskalenko and Suhl \cite{Moskalenko,Suhl} (solid lines at Fig.2). The latter allows to consider the presence of the two observed gaps, $\Delta_L$ and $\Delta_S$, as a bulk property of Gd-1111 compound. In particular, the temperature dependence presented at Fig.2 by open circles is to be attributed to the small bulk (but not surface) gap, which excludes a real-space proximity effect influence \cite{Golubov}. The fact that the $\Delta_{L,S}(T)$ were obtained from namely intrinsic SnS-Andreev spectroscopy supports this conclusion. Besides, the gaps behavior is typical for a two-band superconductor with nonzero interband interaction \cite{Nicol}. At nonzero temperatures, the ``driving'' condensate (with the large gap) induces superconductivity to the second condensate described by the small gap (playing a ``driven'' role) by a k-space (internal) proximity effect \cite{Yanson}. This causes such
features of temperature dependences as bending of $\Delta_L(T)$ and a ``tail'' at $\Delta_S(T)$
driving the small gap to the common $T_C^{local}$. Similar gaps temperature dependence was observed earlier in Mg(Al)B$_2$ superconductor \cite{JETPL04,SSC04,SSC12} and some {\cb other} Fe-based materials \cite{Li,FeSe,FeSeFPS,LOFAFPS}.

\begin{figure}
  \includegraphics[width=0.48\textwidth]{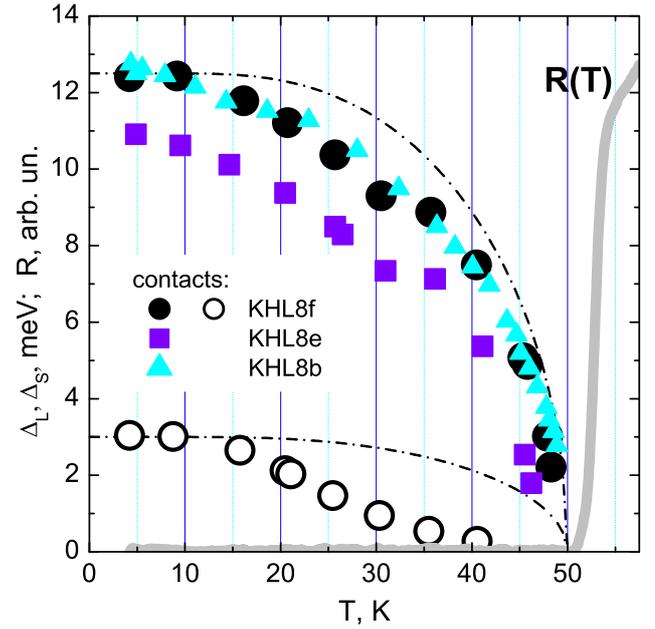}
\caption{Temperature dependence of the large (solid symbols) and the small (open symbols) superconducting gaps for several contacts investigated: sample Khl8, contacts $\sharp f$ (circles, see Fig.2), $\sharp b$ (triangles), and $\sharp e$ (squares). The resistive transition for Khl8 sample (bold line) and single-gap BCS-like functions are also presented.}
\end{figure}

The temperature dependence of superconducting gaps for several studied SnS-arrays
(Khl8 sample, contacts $\sharp f$ (circles, $T_C^{local} \approx 49$\,K, see Fig.~2), $\sharp b$ (triangles, $\Delta_L \approx 12.5$, $T_C^{local} \approx 50$\,K), and $\sharp e$ (squares, $\Delta_L \approx 11$, $T_C^{local} = 48$\,K)) is shown
in Fig.~3. The estimated uncertainty is 10\% for the large and the small gaps. All the experimental $\Delta_L(T)$ dependences are curved in comparison with single-gap BCS-like function. The curvature slightly varies (depending on a particular interband interaction rate), the $\Delta_L(T)$ shape is reproducible on the whole.

Operating with local critical temperature allows to evaluate the BCS-ratio $2\Delta/k_BT_C$ more accurately. The data shown at Fig.3 are $\Delta_L = 11 \div 12.5$\,meV for $T_C^{local} = 48 \div 50$\,K leading to $2\Delta_L/k_BT_C^{local} = 5.3 \div 5.9$. This value exceeds the standard weak-coupling BCS-limit 3.52 thus
pointing at a strong electron-boson coupling in ``driving'' hole bands with the large gap. On the contrary, the small gap ratio is found to be $2\Delta_S/k_BT_C^{local} \ll 3.52$, which supports the induced superconductivity in ``driven'' electron bands within a wide range of temperatures.

To the best of our knowledge, until now there are no experimental data on superconducting gaps value in Gd-oxypnictide, excepting our previous works \cite{Gd,UFN,GdJPCS}. As was shown theoretically in \cite{Kuchinskii}, the BCS-ratio is nearly independent on $T_C$ for oxypnictides. The latter enables a comparison between $2\Delta_L/k_BT_C$ for 1111-materials with different $T_C$ (for a review, see Table 1 in \cite{Gd}, and Table 2 in \cite{Seidel}). In this respect, our $2\Delta_L/k_BT_C$ results agree with point-contact Andreev reflection (PCAR) spectroscopy data from \cite{Samuely,Yates} and IMARE measurements \cite{LOFA,LOFAFPS}.

In conclusion, oxygen-deficient GdFeAsO$_{0.88}$ samples were studied by intrinsic multiple Andreev reflections effect (IMARE) spectroscopy (realized by the ``break-junction'' technique). We found two superconducting gap values: the large gap $\Delta_L = 11.8 \pm 1.2$\,meV with the BCS-ratio $2\Delta_L/k_BT_C = 5.3 \div 5.9$ and the small gap $\Delta_S = 3 \pm 0.3$\,meV with $2\Delta_S/k_BT_C \ll 3.52$ for local critical temperature $T_C^{local} = 48 \div 50$\,K. The obtained  temperature dependences of the large and the small superconducting gaps exhibit distinct shape, both deviating from the single-band BCS-like function. At the same time, the $\Delta_{L,S}(T)$ courses agree with the two-band BCS-like model by Moskalenko and Suhl \cite{Moskalenko,Suhl}, which confirms the presence of two distinct superconducting gaps as a bulk property of GdFeAsO$_{0.88}$.

\begin{acknowledgements}
This work was supported by Russian Foundation for Basic Research, Programs of Russian
Academy of Sciences, Russian Ministry of Education and Sciences, and was performed
using facilities of the Shared Research Facilities Center at LPI. We thank I.V. Morozov and D. Daghero for fruitful discussions.
\end{acknowledgements}

\end{document}